\documentclass[aps,prl,reprint,superscriptaddress,nofootinbib,longbibliography]{revtex4-2}

\usepackage{amsmath,amssymb,mathtools}
\usepackage{bm}
\usepackage{graphicx}
\usepackage{booktabs}
\usepackage{xcolor}
\usepackage[colorlinks=true,citecolor=blue,linkcolor=blue,urlcolor=blue]{hyperref}
\usepackage{makecell}

\newcommand{\ii}{\mathrm{i}}
\newcommand{\dd}{\mathrm{d}}
\newcommand{\Tr}{\operatorname{Tr}}
\newcommand{\norm}[1]{\left\lVert #1\right\rVert}
\newcommand{\abs}[1]{\left\lvert #1\right\rvert}

\newcommand{\re}{\operatorname{Re}}

\begin{document}

\title{Second-Chern Bounds in Non-Abelian Quantum Geometry}

\author{Junwen Zhao}
\thanks{These two authors contributed equally to this work.}
\affiliation{Fudan University, Shanghai 200433, China}
\affiliation{New Cornerstone Science Laboratory, Department of Physics, School of Science, Westlake University, Hangzhou 310024, Zhejiang, China}
\author{Zhiming Pan}
\thanks{These two authors contributed equally to this work.}
\affiliation{Department of Physics, Xiamen University, Xiamen 361005, China}
\author{Kang Yang}
\email{yangkang@westlake.edu.cn}
\affiliation{Department of Physics, School of Science, Westlake University, Hangzhou 310024, Zhejiang, China}
\author{Congjun Wu}
\email{wucongjun@westlake.edu.cn}
\affiliation{New Cornerstone Science Laboratory, Department of Physics, School of Science, Westlake University, Hangzhou 310024, Zhejiang, China}
\affiliation{Institute for Theoretical Sciences, Westlake University, Hangzhou 310024, Zhejiang, China}
\affiliation{Key Laboratory for Quantum Materials of Zhejiang Province, School of Science, Westlake University, Hangzhou 310024, Zhejiang, China}
\affiliation{Institute of Natural Sciences, Westlake Institute for Advanced Study, Hangzhou 310024, Zhejiang, China}

\date{\today}
\begin{abstract}
We study the quantum geometry of doubly degenerate energy levels in a four-dimensional parameter space.  
We find that the scalar quantum metric $g$ and the Berry curvature $F$ obey $(\operatorname{tr} g)^2/16\geq\sqrt{\det g}\geq |2\Tr(F\wedge F)-\Tr F\wedge \Tr F|/24$. 
The first inequality characterizes the anisotropy in the metric. 
The second determinant inequality measures the self-duality of the traceless $SU(2)$ part of the curvature under Hodge star operation and the algebraic closedness of the inter-level polarization amplitudes under $SU(2)$ rotations in the doubly degenerate levels.  
The saturation of the determinant bound imposes a quaternionic Cauchy-Riemann equation, analogous to the complex analyticity imposed by ideal-band conditions in two-dimensional Chern insulators. 
As examples, four-band Dirac Hamiltonians automatically saturate the determinant bound and possess a topological zero in $\operatorname{tr}(F\wedge F)$. 
We compare the differences between Kramers degeneracy and ordinary $U(2)$ degeneracy. In addition to the non-Abelian geometric bound, the latter also obeys an independent first-Chern bound.
\end{abstract}

\maketitle

\textit{Introduction.---}
Quantum metric and Berry curvature describe how quantum states vary with respect to parameters. 
They correspond to the symmetric and antisymmetric parts of the quantum geometric tensor (QGT)~\cite{provost1980riemannian,berry1984quantal,anandan1990geometry,xiao2010berry}.  
In two dimensions, the integral of Berry curvature gives the first Chern number, which determines the quantized Hall response and obstructs exponentially localized Wannier functions~\cite{thouless1982quantized,haldane1988model,brouder2007exp}. Meanwhile, quantum metric quantitatively characterizes the localization length of electrons and optical conductivity~\cite{marzari1997maximally,resta1999,PhysRevB.62.1666}. 
It can contribute to the superfluid weight and support superconductivity even in a flat band~\cite{peotta2015sf,hazra2019bounds,torma2023essay}. The quantum metric is known to be bounded by the Berry curvature \cite{roy2014band,mera2021kahler}. This imposes fundamental relations among various physical phenomena \cite{PhysRevB.95.024515,PhysRevX.14.011052}.
Landau levels provide basic ideal examples where the metric and Berry curvature are exactly locked to each other~\cite{roy2014band,mera2021kahler}.  
This structure has guided the construction of ideal Chern bands and fractional Chern insulators~\cite{claassen2015position,ledwith2020fractional,PhysRevLett.127.246403,PhysRevResearch.5.L012015}.

The geometric bound between the quantum metric and Berry curvature originates from the semi-positivity of the Abelian QGT~\cite{ma2010abelian,roy2014band,mera2021kahler}.  
Saturation of the bound implies a reciprocal-space Cauchy-Riemann equation for the band structure \cite{claassen2015position,mera2021kahler}. 
This exactly corresponds to the descriptions of Landau levels and moir\'e graphene in the chiral limit~\cite{PhysRevLett.122.106405,ledwith2020fractional,PhysRevLett.127.246403}. Intuitively, the complex structure imposed by the Cauchy-Riemann equation fixes the Abelian Berry curvature to the volume determined by the metric.

The geometry changes qualitatively for degenerate multi-band structures.  
Adiabatic transport in the parameter space rotates the states within the degenerate occupied subspace, 
and the QGT becomes matrix-valued~\cite{wilczek1984app}. This leads to intrinsic non-Abelian quantum geometry whose components cannot be fixed by complex analytic equations.
An example is the doubly-degenerate energy level with a $SU(2)$ gauge structures.
In four-dimensional (4D) parameter space, the $SU(2)$ gauge structure produces the topological second Chern number that probes matrix rotations in the degenerate doublet. 
Its basic realization is the Yang monopole, and it underlies the four-dimensional quantum Hall effect and higher-dimensional $SU(2)$ Landau levels~\cite{yang1978gen,zhang2001four,li2013high,li2013topo,kolodrubetz2016measuring,sugawa2018second,zhao2026incomp,hatsugai2010quat}.
The non-Abelian QGTs and their relation to topological invariants have attracted interests recently~\cite{ding2022extracting,mera2022relating}.

In this Letter, we consider a gapped and doubly degenerate band structure and derive geometric bounds for the quantum metric.  
From Hodge decomposition, we obtain the bounds relating the quantum metric to the second Chern class.  
Their saturation induces three Cauchy-Riemann equations forming a quaternion algebra.
Minimal four-band Dirac models saturate the determinant bound while exhibit singular points in the geometry.
The results apply to both $SU(2)$ Kramers pairs and ordinary $U(2)$ degeneracies, while the latter is found to satisfy two independent sets of quantum geometric inequalities.

\textit{$SU(2)$ geometry of a degenerate doublet.---}
Consider a smooth $N$-band Hamiltonian $H(\bm\lambda)$ in a 4D $\bm{\lambda}$-space $M$.  
Two degenerate occupied bands $|u_n\rangle$ are gapped from other empty bands.
We collect these two states as a local frame as $\Psi=(|u_1\rangle,|u_2\rangle)$ and introduce the projector $P$ and the matrix $X_{\mu}$,
\begin{align}
P=\Psi\Psi^{\dagger},\qquad
X_{\mu}\equiv (1-P) \partial_{\mu}\Psi\equiv D_\mu\Psi.
\end{align}
Here we abbreviate $\partial_\mu\equiv\partial_{\lambda^\mu}$ and introduce the derivative $D_\mu$ tangent to $\Psi$. The matrix $X_\mu$ describes the virtual transitions to empty eigenstates $|u_\ell\rangle$:
$(X_{\mu})_{\ell n}\equiv 
 \langle u_\ell|\partial_\mu u_n\rangle
 =\langle u_\ell|\partial_\mu H|u_n\rangle/
 (\varepsilon_n-\varepsilon_\ell).$
This is the standard interband polarization amplitude used in band geometry and response theory~\cite{xiao2010berry,xu2018quantized,mera2022degenerate,liu2025quantum,verma2026quantum}.
The non-Abelian Berry connection $[A_{\mu}]_{mn}=A_{\mu}^{mn} \equiv \ii\langle u_m|\partial_\mu u_n\rangle$ describes the variation of the local frame.

The non-Abelian QGT is given by the Gram matrix:
\begin{equation}
Q_{\mu\nu} =X_{\mu}^{\dagger} X_{\nu}
=G_{\mu\nu} -\frac{\ii}{2}F_{\mu\nu},
\label{eq:definition}
\end{equation}
whose symmetric part gives the quantum metric $G_{\mu\nu}=\frac{1}{2} \big(Q_{\mu\nu} +Q_{\nu\mu} \big)$ and the antisymmetric part gives the non-Abelian Berry curvature $F_{\mu\nu}=\ii \big(Q_{\mu\nu} -Q_{\nu\mu} \big)$.
The component-wise expressions are explicitly given by $[Q_{\mu\nu}]_{nm}=\langle \partial_{\mu} u_n| (1-P) |\partial_{\nu} u_m\rangle$.
With the Berry connection $A_{\mu}$,
the curvature $F_{\mu\nu}$ can be formulated in the conventional non-Abelian curvature form,
\begin{equation}
F_{\mu\nu}=\partial_\mu A_\nu-\partial_\nu A_\mu -\ii[A_\mu,A_\nu]
\label{eqS:curvature-response}
\end{equation}
From hereby, we introduce $F=\sum_{\mu,\nu}F_{\mu\nu}\,\mathrm{d}\lambda^\mu\wedge\mathrm{d}\lambda^\nu/2$.
We denote the scalar quantum metric as
\begin{equation}
g_{\mu\nu}=\Tr G_{\mu\nu}
=\operatorname{Re}\Tr(X_\mu^\dagger X_\nu).
\label{eq:scalar-metric-main}
\end{equation}
Here ``$\Tr$'' traces over the occupied-band indices, summing the transition probability from two occupied bands to all empty bands.
It measures the total probability transferred out of the occupied doublet by an infinitesimal change of parameters.

As $\bm{\lambda}$ varies, the frame $\Psi$ rotates. When $\bm{\lambda}$ is scanned over a loop in the parameter space $M$, the parallel-transported frame changes by a $U(2)$ rotation $\Psi'=\Psi U$. A $U(2)$ transformation can be decomposed into an Abelian phase times a $SU(2)$ non-Abelian component. As all non-Abelian information is encoded in the $SU(2)$ subgroup, we first focus on situations where $U\in SU(2)$. This is realized through two Kramers-degenerate levels $|u_2\rangle=\Theta |u_1\rangle$ related by an antiunitary symmetry $| \Theta^2=-1$, while the more general $U(2)$ cases will turn out to be just enriched with Abelian quantum geometry relations. Under this assumption, $\Tr F=0$, i.e., no net phase is produced in parallel transports of $\Psi$ along any loops in $M$. 
As $SU(2)\simeq Sp(1)$, a quaternion structure appears in the Hilbert spaces spanned by $\Psi$. 
With this identification, the negative-energy doublet of four-band Dirac Hamiltonians resembles the canonical $\mathbb HP^1$ Yang monopole \cite{yang1978gen}.

\textit{Bounds of the second Chern class.---}
For the $SU(2)$ case, the curvature can be decomposed into
\begin{align}
F_{\mu\nu}=\sum_a F_{\mu\nu}^a\sigma_a,\quad
F^a=\frac{1}{2}\sum_{\mu,\nu}F_{\mu\nu}^a \,\mathrm{d}\lambda^\mu\wedge\mathrm{d}\lambda^\nu
\end{align}
where the Pauli matrices $\sigma_a$ ($a=1,2,3$) act on the two-dimensional band index.
Changing the local basis of the two occupied bands rotates the three $\sigma_a$ components among themselves.
For traceless curvature $F$,
the second Chern class (density) $c_2$ is reduced to
\begin{align}
c_2&=\frac{1}{8\pi^2}\Tr (F\wedge F)
=\frac{1}{4\pi^2}\sum_a F^a\wedge F^a.
\label{eq:main-c2-math}
\end{align}
Its integrals give the second Chern number $C_2=\int c_2\in\mathbb Z$. Since $c_2$ is a four-form proportional to the volume form, we parameterize $c_2(\bm{\lambda})=s(\bm{\lambda})|c_2(\bm{\lambda})|\dd^4\lambda$, where $s(\bm{\lambda})=\pm1$ is the sign of $c_2$ relative to the volume form.

The main result of this work is the following inequalities,
\begin{equation}
\frac{1}{16} \big(\operatorname{tr} g\big)^2 
\geq \sqrt{\det g}
\geq \frac{1}{6}\left|\sum_a\Tr (F^a\wedge F^a)\right|.
\label{eq:two-bounds-main}
\end{equation}
Here ``tr'' is for parameter-space indices $\operatorname{tr}g=\sum_{\mu}g_{\mu\mu}$.
The first inequality arises from the inequality between the arithmetic mean and the geometric mean of the eigenvalues of $g$. It is saturated when the metric is isotropic $g\sim 1_{4\times 4}$. The determinant bound can be saturated for anisotropic metric.
These inequalities relate three levels of information in the interband polarization amplitudes.
The trace of the quantum metric $g_{\mu\nu}$ measures the total strength of the deformation of the occupied space, 
whereas $\sqrt{\det g}$ measures the four-dimensional volume spanned by four independent deformations.  
The second Chern class retains only the $SU(2)$ geometric part of this volume.  
For a closed parameter space $M$, the integral of Eq.~\eqref{eq:two-bounds-main} leads to a topological bound
\begin{equation}
\int\frac{d^4\lambda}{16} \big(\operatorname{tr} g\big)^2\ge\int \sqrt{\det g}\ d^4\lambda
\geq \frac{2\pi^2}{3}|C_2|.
\end{equation}
The bound may thus be viewed as the minimal quantum-geometric cost of carrying a quantized $SU(2)$ instanton charge.
In the following, we give a geometric proof for Eq.~\eqref{eq:two-bounds-main} and explain the saturation conditions.

\textit{Hodge decomposition of the curvature.---}
At every point where $g$ is nondegenerate ($\det g>0$),  the metric defines a Hodge star operation $*_g$, maping two-forms to two-forms.
The norm convention for a given two-form $K$ is defined as $\norm{K}_g^2=\sum_{\mu,\nu,\rho,\sigma}K_{\mu\nu}K_{\rho\sigma}g^{\mu\rho}g^{\nu\sigma}/2\ge 0$, where $g^{\mu\nu}\equiv[g^{-1}]^{\mu\nu}$.
The two-form $F^a$ can be decomposed into self-dual and anti-self-dual sectors,
\begin{equation}
F_{\pm}^{a}=\frac12(F^a\pm *_gF^a),\qquad
*_gF_{\pm}^a=\pm F_{\pm}^a.
\label{eq:hodge-main}
\end{equation}
This decomposition reflects the structure $SO(4)\simeq {SU(2)_{+}\times SU(2)_{-}}/{\mathbb Z_{2}}$.
Geometrically, Hodge star simply pairs complementary 2D planes, $(12)\leftrightarrow(34)$, $(13)\leftrightarrow(42)$, and $(14)\leftrightarrow(23)$. The parity under Hodge star operation generalizes the two-dimensional handedness/chirality to four dimensions.

The Hodge decomposition reorganizes the Chern class into
\begin{equation}
c_2=\frac{1}{4\pi^2}
\left(\sum_a\norm{F_+^a}_g^2-\sum_a\norm{F_-^a}_g^2\right)\sqrt{\det g}\,\dd^4\lambda.
\label{eq:c2-hodge-main}
\end{equation}
This expression gives $c_2$ an interpretation as the difference between the two handed Hodge sectors.
The value of $|c_2|$ is maximized when all $F^a$ belong to the same Hodge sector corresponding to the sign of $c_2$.
With the decomposition Eq.~\eqref{eq:c2-hodge-main}, we write
the difference between the determinant and the Chern class as
\begin{align}
\Delta=\sqrt{\det g} -\frac{2\pi^2}{3} |c_2|
=\frac{1}{6} (\Delta_Q+\Delta_H)\sqrt{\det g},
\end{align}
where the two contributions are given by
$\Delta_H= 2\sum_a\norm{F_{-s}^{a}}_g^2,
\Delta_Q= 6-\sum_a\norm{F^a}_g^2. $
We will show that both contributions are non-negative $\Delta_H\ge 0, \Delta_Q\ge 0$. The determinant inequality in Eq.~\eqref{eq:two-bounds-main} follows. The saturation is reached when $\Delta_H=\Delta_Q=0$.

The Hodge contribution $\Delta_H$ provides a local measure of the deviation from the Bogomolny self-duality condition in four-dimensional Yang-Mills theory \cite{belavin1975pseudo,Bogomolny1976},
which is explicitly non-negative.
It is the Berry curvature component whose handedness is opposite to the local sign of $c_2$.
It is saturated when
\begin{equation}
 *_gF^a=sF^a\ (a=1,2,3)\Leftrightarrow \Delta_H=0.
 \label{eq:common-chirality}
\end{equation}
The $\Delta_Q$ part measures the failure of the four-dimensional tangent space of $M$ to preserve quaternion rotations. 
To demonstrate $\Delta_Q\ge 0$, we introduce the $(1,1)$-form $[J^a]^{\mu}_\nu=\sum_\rho g^{\mu\rho}F^a_{\rho\nu}$ and its conjugate $[(J^a)^\dagger]^\mu_\nu=\sum_{\rho,\sigma}g^{\mu\rho}[J^a]^\sigma_\rho g_{\sigma\nu}$, where $g^{\mu\nu}$ is the inverse of $g$ as a $4\times 4$ matrix.
We have the relation $\norm{F^a}_g^2=\operatorname{tr}[(J^a)^\dagger J^a]/2$. 
We treat the QGT as a quaternion-valued bilinear form in the tangent space of $M$, where $g$ is the scalar component and $F^a$ are the coefficients of the three quaternions \cite{supple}. 
The action of $Q$ on four dimensional tangent vectors $v,v'\in TM$ is done by contracting the $\mu,\nu$ indices of $v,v'$. 
The quaternion Cauchy-Schwarz inequality imposes $|Q(v,v')|_{\mathbb H}\le \sqrt{g(v,v)g(v',v')}$. 
On the other hand, by definition, we have $g(v,J^av')=F^a(v,v')$. As $F^a(v,v')$ is a summand in the quaternion $Q(v,v')$, we have $|g(v,J^av')|\le \sqrt{g(v,v)g(v',v')}$. 
Taking $v=J^av'$, this shows that $g(J^av',J^av')\le g(v',v')$. 
The $J^a$ operation makes the vector lengths smaller. 
This demonstrates $||(J^a)^\dagger J^a||\le 1$ (operator norm). 
Tracing over all spatial indices, we have $\norm{F^a}_g^2=\operatorname{tr}[(J^a)^\dagger J^a]/2\le 2$. 
Summing over all $a=1,2,3$, this yields $\Delta_Q\ge 0$.

\textit{Quaternion Cauchy-Riemann equations.---} We investigate when $\Delta_Q=0$. Together with Eq.~\eqref{eq:common-chirality}, the saturation condition for the determinant bound is obtained. According to the discussion, this requires $\operatorname{tr}[1-(J^a)^\dagger J^a]=0$. As $1-(J^a)^\dagger J^a$ is semi-positive definite, we have $(J^a)^\dagger J^a=1$. Since $(J^a)^\dagger=-J^a$, this requires $[J^a]^2=-1$.

Putting this condition back to the quantum metric, we have $g\big(v,(1-[J^a]^\dagger J)v\big)=0$ for any vector $v$. As $g(v,v')$ originates from the inner products of $D_v\Psi$ [c.f. Eq.~\eqref{eq:scalar-metric-main}], this condition tells that certain combinations of $D_v\Psi$ are null vectors. By analyzing these null relations \cite{supple}, we find the following conditions at $\det g(\bm{\lambda)}>0$ points: 
\begin{equation}
    D_\mu\Psi(-i\sigma_a)+\sum_{\nu}[J^a]^\nu_\mu D_\nu \Psi=0\Leftrightarrow \Delta_Q=0,\label{eq_qcr}
\end{equation}
where $J^a(\bm\lambda)$ satisfies the quaternion algebra: $J^1J^2=J^3$ and $(J^a)^2=-1_{4\times 4}$. The equation is covariant under diffeomorphism transformations of $M$ and $SU(2)$ rotations in $\Psi$. It tells that a quaternion action $(-i\sigma_a)$ on the doublet isospin space commutes with a quaternion action $J^a$ in the tangent space of the manifold $M$. This is a triple Cauchy-Riemann equation. To see this, in two dimensions, if we set $J_\alpha^\beta=\epsilon^{\beta\alpha}$ where $\alpha,\beta=1,2$, the Cauchy-Riemann equation takes the form:
\begin{equation}
    \partial_\alpha|\psi\rangle i+\sum_{\beta}J^\beta_\alpha \partial_\beta |\psi\rangle=0, \ \textrm{2D Cauchy Riemann}.
\end{equation}
Comparing the two equations, Eq.~\eqref{eq_qcr} corresponds exactly to three quaternion Cauchy-Riemann equations related by $SO(3)$ quaternion rotations. It is the four-dimensional analog of closedness under complex actions $J^2=-1$ for ideal Abelian Chern bands~\cite{ledwith2020fractional,PhysRevLett.127.246403}.

The preceding demonstration proves $\Delta(\bm{\lambda})\ge 0$ and its saturation condition for $\det g(\bm{\lambda})>0$ points. 
When $\det g(\bm{\lambda})=0$, $X_\mu$ are linear-dependent. 
The vector space spanned by them has dimensions lower than four. 
All four forms on this space have to vanish and so does $c_2$, which is an exterior power based on $X_\mu$.
The determinant inequality of Eq.~\eqref{eq:two-bounds-main} simply becomes 0 on both sides.

\textit{Four-band Hamiltonians.---}
We demonstrate two facts in four-band Dirac Hamiltonians: (1) All such four-band models saturate the determinant bound; (2) For $M$ being a 4D torus, there exists $\bm{\lambda}$ such that $\det g(\bm{\lambda})=\operatorname{tr}[F(\bm{\lambda})\wedge F(\bm{\lambda})]=0$. It is impossible to build a topological four-band model with a constant second Chern density.

After neglecting an overall constant, four-band Dirac Hamiltonians are parametrized by five real coefficients $\textrm{d}_A\in \mathbb R$ in front of $4\times 4$ Dirac Gamma matrices $\Gamma_A$~\cite{yang1978gen,zhang2001four,xiaoliang2008}
\begin{equation}
H(\bm\lambda)=\sum^5_{A=1} \textrm{d}_A(\bm\lambda)\Gamma_A,
\label{eq:dirac-main}
\end{equation}
The four bands are organized into two doubly degenerate pairs. The projector to the two negative-energy bands is uniquely expressed by the normalized vector $\bm n$: $P=(I_4-\sum_A n_A\Gamma_A)/2$ with $n_A=\textrm{d}_A/|\mathrm{\textbf{d}}|$. Substituting the projector into Eq.~\eqref{eq:definition}, the quantum geometry is given by
\begin{equation}
g_{\mu\nu}=\frac12\partial_\mu{\bm n}\cdot
 \partial_\nu\bm n,\qquad \Tr(F\wedge F)=\frac{1}{8}\omega_{S^4}.
\label{eq:dirac-identity-main}
\end{equation}
where $\omega_{S^4}$ is the infinitesimal area swept by the $\bm n$ vectors: $\omega_{S^4}=\sum_{ABCDE}\epsilon_{ABCDE}n_A (dn_B) \wedge (dn_C) \wedge (dn_D) \wedge (dn_E)$ \cite{xiaoliang2008,mera2022relating}.
The orthogonal projector $P$ to the sub-Hilbert space spanned by $|u_1\rangle,|u_2\rangle$ is uniquely characterized by the vector $\bm n\in S^4$. As a result, the quantum geometry is given by the geometry of $S^4$, which is the quaternion projective space $\mathbb HP^1$. This is the four-band analog of the two-band Bloch sphere $\mathbb CP^1\simeq S^2$. 

Now we prove the two statements. The derivative $\partial_\mu\bm n$ has the geometric meaning of how $\bm n$ moves over $S^4$ as $\bm \lambda$ varies. The metric $g$ is the Gram matrix for the corresponding $d\bm n$. The Gram-determinant identity indicates that $\det g$ is proportional to the norm square of $\prod\wedge d\bm n$ projected in the tangent space of $S^4$. This is exactly $\Tr(F\wedge F)$. The determinant bound is thus always saturated.
Assume that $\det g> 0$ for all $\bm{\lambda}$, this tells the tangent map $dP$ at every $\bm{\lambda}$ is an isomorphism, leading to a local diffeomorphism. A local diffeomorphism between two compact spaces is a covering map. This would map the fundamental group $\pi_1(T^4)=(\mathbb Z)^4$ injectively into $\pi_1(S^4)=0$, which is impossible. Thus, there must be $\bm{\lambda}$ where $\det g(\bm{\lambda})=\Tr F(\bm{\lambda})\wedge F(\bm{\lambda})=0$.

These two properties are reminiscent of the previous results on 2D two-band models. In 2D two-band Hamiltonians, it is known that the Abelian Berry curvature is always saturates the determinant bound $\sqrt{\det g}=|F|/2$, and there exist points in the Brillouin zone where the Berry curvature vanishes \cite{mera2021kahler,SciPostPhys.12.4.118}. This obstructs the possibility of realizing constant Berry curvature in two-band topological models, a condition suggested to favor fractional Chern insulators \cite{goerbig2012,parameswaran2013fqh}. The statements presented here are the 4D counterparts.

\textit{Situations for $U(2)$ doublet.---}
In general, a degenerate doublet $\Psi$ may transform by a $U(2)$ rotation when parallel transported along a parameter loop. The Abelian Berry curvature $\Tr F$ is non-vanishing, giving rise to nontrivial first Chern classes $c_1=\Tr F/(2\pi)$.
Meanwhile, the second Chern class receives one more contribution \cite{tu2017differential}: 
$c_2=\left[
\Tr(F\wedge F)-(\Tr F)\wedge(\Tr F)\right]/(8\pi^2)$. The geometric inequality Eq.~\eqref{eq:two-bounds-main} still holds at every $\bm{\lambda}$, and can be rewritten as~\cite{supple}
\begin{equation}
U(2)\textrm{ doublets}: \sqrt{\det g}\ge  \frac{2\pi^2}{3}
\left| c_2+\frac14c_1\wedge c_1 \right|.\label{eq_u2b}
\end{equation}
This equation is also saturated when both Eqs.~\eqref{eq:common-chirality} and \eqref{eq_qcr} hold \cite{supple}. 
We can further integrate both sides to yield a global topological bound:
\begin{equation}
\int \sqrt{\det g}\ d^4\lambda  \geq \frac{2\pi^2}{3}
\left|C_2+\frac14C_{1,1}\right|,
\end{equation}
where we use the absolute value inequality for integrals.

On the other hand, the Abelian component of $U(2)$ provides an additional quantum geometric inequality. Besides the second Chern number, there is another 4D Chern number $C_{1,1}=\int c_1\wedge c_1 \in\mathbb Z$ \cite{milnor1974characteristic}. 
The Wirtinger inequality asserts \cite{ballmann2006lectures,HASHIMOTO2026105969}:
\begin{equation}
\int \sqrt{\det g}\ d^4\lambda\geq
\Big|\int \frac18(\Tr F)\wedge(\Tr F)\Big|
=\frac{\pi^2}{2} |C_{1,1}|.
\label{eq:u2-c1-main}
\end{equation}
This is a 4D topological bound in the first Chern class. Therefore, there are two independent bounds Eqs.\eqref{eq_u2b} and \eqref{eq:u2-c1-main} on the quantum geometry of $U(2)$ doubly degenerate bands. This difference from $SU(2)$ doublets can be explained by the following geometric distinction. For $SU(2)\simeq Sp(1)$, the Hilbert space spanned by the doublet corresponds to the quaternion projective space $\mathbb HP^{N/2-1}$, which is a quaternionic K\"ahler manifold imposing the bound Eq.~\eqref{eq:two-bounds-main}. While for $U(2)$, the Hilbert space spanned by the doublet corresponds to the complex Grassmannian $\textrm{Gr}_2(\mathbb C^N)$, which is both K\"ahler and quaternionic K\"ahler \cite{wolfSpace,salamon1982quaternionic}. This explains why $U(2)$ doublets have two different quantum geometry inequalities.

A minimal saturation $U(2)$ example is a three-level Hamiltonian on $M=\mathbb CP^2$, locally parameterized by ~\cite{liu2007topo},
\begin{align}
H=2|v\rangle\langle v|-I_3,\quad
|v\rangle&=
\frac{(1,w_1,w_2)^T}{\sqrt{1+\abs{w_1}^2+\abs{w_2}^2}},
\label{eq:cp2-main}
\end{align}
The two complex numbers $w_1,w_2$ give four real parameters.
The state $|v\rangle$ is the empty level at energy $+1$, while the orthogonal two-dimensional subspace consists of two degenerate occupied levels at energy $-1$.
The global parameter space of the normalized three-component vector modulo its phase is denoted $\mathbb C P^2$.
All $(X_{\mu})_{\ell n}$ amplitudes pass to the single empty level, and Eq.~\eqref{eq:u2-c1-main} is saturated locally~\cite{supple}:
\begin{equation}
 c_2=-c_1\wedge c_1,\quad
 \sqrt{\det g}=\frac{\pi^2}{2}\abs{c_1^2}
 =\frac{\pi^2}{2}\abs{c_2}.
 \label{eq:cp2-saturation-main}
\end{equation}
Interestingly, the same model also saturates the $U(2)$ bound in
Eq.~\eqref{eq_u2b}.

\textit{Discussion.---}
The geometric bounds place constraints on non-Abelian quantum geometry, which characterizes interband polarization amplitudes.
It does not assume flat bands, weak coupling, or a particular lattice model. Unlike the two dimensional geometric bound whose saturation selects a chirality in the wavefunctions, saturation of the four dimensional bounds selects a Hodge sector.
The metric must be fully converted into a single-handed set of relative $SU(2)$ rotations without spectator transition channel.

This aspect is useful for experiments and numerics.
Quantum metrics can be extracted from excitation rates under weak parameter modulation \cite{kim2025direct,kang2025measurements}, while non-Abelian curvature can be reconstructed from state tomography or Wilson-loop protocols~\cite{kolodrubetz2016measuring,ding2022extracting,sugawa2018second}.
The two contributions in the determinant inequality correspond to different physical processes. The contribution
$\Delta_H$ detects cancellation between complementary parameter loops.
In comparison, the contribution $\Delta_Q$ detects leakage into non-adiabatic interband channels.
The latter should be especially sensitive to remote-band mixing.

The determinant bound can be saturated by an anisotropic QGT, while the trace bound requires an isotropy. This might lead to differences when the geometry is defined with respect to pumping parameters $\bm\lambda=(k_x,k_y,k_z,\phi)$. For example. the relative weight assigned to the phase $\phi$ must be fixed before the metric trace has a physical meaning.
Note that saturation of the bounds at every parameter point does not immediately yield the topological bound saturation.
Local geometric bounds test the closedness of virtual interband transitions at each parameter.
Saturation of the topological bound further requires a fixed sign of the second-Chern density.
Neither condition by itself implies a flat dispersion or a correlated many-body phase.
They instead identify the band geometry that minimizes the wave-function volume needed to carry a given $SU(2)$ topological density.
This provides a starting point for extending ideal-band ideas from Chern bands to four-parameter non-Abelian systems.

\textit{Note added.---}
While this manuscript was being prepared, Ref.~\cite{lim2026quat} on quaternionic K\"ahler quantum geometry for Kramers pairs appeared,.  
That work derives an equivalent second-Chern quantum-volume bound and characterizes its saturation through pushforward tangent space properties,
including the saturation of minimal quaternionic four-band models. 
The present work derives the gap in the quantum geometric bound from the Hodge decomposition and considers general $U(2)$ cases without Kramers degeneracy.
We have also learned that G.~Y.~Cho et al are independently studying related aspects of non-Abelian quantum geometry~\cite{cho2026inprep} and thank them for sharing a preliminary manuscript.

\textit{Acknowledgments.---}
C.W. is supported by the National Natural Science Foundation of China under Grants No. 12234016 and No. 12174317.  
Z.P. is supported by the National Natural Science Foundation of China under Grant No. 12504219. KY thanks Bruno Mera for pointing out the Wirtinger inequality and is supported by a start-up grant from Westlake University.
This work has been supported by the New Cornerstone Science Foundation.

\bibliography{references}

\clearpage
\onecolumngrid
\appendix

\onecolumngrid

\title{Supplemental Material for Second-Chern Bounds in Non-Abelian Quantum Geometry}

\author{Junwen Zhao}
\thanks{These two authors contributed equally to this work.}
\affiliation{Fudan University, Shanghai 200433, China}
\affiliation{New Cornerstone Science Laboratory, Department of Physics, School of Science, Westlake University, Hangzhou 310024, Zhejiang, China}
\author{Zhiming Pan}
\thanks{These two authors contributed equally to this work.}
\affiliation{Department of Physics, Xiamen University, Xiamen 361005, China}
\author{Kang Yang}
\email{yangkang@westlake.edu.cn}
\affiliation{Department of Physics, School of Science, Westlake University, Hangzhou 310024, Zhejiang, China}
\author{Congjun Wu}
\email{wucongjun@westlake.edu.cn}
\affiliation{New Cornerstone Science Laboratory, Department of Physics, School of Science, Westlake University, Hangzhou 310024, Zhejiang, China}
\affiliation{Institute for Theoretical Sciences, Westlake University, Hangzhou 310024, Zhejiang, China}
\affiliation{Key Laboratory for Quantum Materials of Zhejiang Province, School of Science, Westlake University, Hangzhou 310024, Zhejiang, China}
\affiliation{Institute of Natural Sciences, Westlake Institute for Advanced Study, Hangzhou 310024, Zhejiang, China}

\date{\today}

\maketitle

\onecolumngrid

\section{Quaternion Cauchy-Schwarz inequality and the saturation condition}

We demonstrate the quaternion Cauchy-Schwarz inequality used in the main text. 
We start from a system with a $\Theta^2=-1$ symmetry. 
This leads to the essential steps in the geometric bound. 
We find that the key ingredient is a local quaternion structure in the wavefunctions at each $\bm\lambda$. 
The local quaternion structure can exist even for a $U(2)$ doublet without $\Theta^2=-1$ symmetries. 
Therefore, the $\Theta^2=-1$ condition can be removed and a more general proof that further applies to $U(2)$ cases will be given in the next section.

We first show that the Hilbert space of a $N$-level system with a $\Theta^2=-1$ symmetry can be identified as a $N/2$-dimensional quaternion space. 
This is done by constructing an orthonormal basis of it by induction. 
First, we pick a state $|e_1\rangle$. 
Due to Kramers degeneracy, the state $\Theta|e_1\rangle$ is a state orthonormal to $|e_1\rangle$. 
The doublet $|e_1\rangle,\Theta|e_1\rangle$ constitute the first two vectors in the orthogonal basis. 
Then we look at the $(N-2)$-dimensional plane perpendicular to $\{|e_1\rangle,\Theta|e_1\rangle\}$ and pick a state $|e_2\rangle$ in it. 
As before, $\Theta|e_2\rangle$ is orthonormal to $|e_2\rangle$ as well as $|e_1\rangle,\Theta|e_1\rangle$. 
By repeating this process, we obtain a basis $|e_1\rangle,\Theta|e_1\rangle,\dots,|e_{N/2}\rangle,\Theta|e_{N/2}\rangle$.

\subsection{Proof of the inequality}
The two-dimensional Hilbert space spanned by each Kramers pair is a one-dimensional quaternion space. 
Taking $|e_1\rangle,\Theta|e_1\rangle$ for example, 
any state in the sub-Hilbert space spanned by the Kramers pair is written as $(a+ib)|e_1\rangle+(c+id)\Theta|e_1\rangle$, 
where $a,b,c,d$ are real numbers. 
Recombining the coefficients, we have $(a+bi+c\Theta+di\Theta)|e_1\rangle$. 
Note that the three operations $i,\Theta,i\Theta$ satisfy the quaternion algebra. 
We can identify the coefficients as a quaternion number. 
By applying this procedure to all $|e_j\rangle$, 
we have the expansion for any state $|\psi\rangle$
\begin{equation}
|\psi\rangle=\sum_j q_j|e_j\rangle, \qquad 
q_j=a_j+b_j i+c_j \Theta+d_ji\Theta.
\end{equation}
Thus, the Hilbert space $\mathbb C^N$ is identified with a quaternion vector space $\mathbb H^{N/2}$. 
Every Kramers pair forms a $Sp(1)\simeq SU(2)$ doublet. 
Every state can be represented by a $N/2$-dimensional quaternion vector $q=(q_1,q_2,\dots,q_{N/2})^T$.

With this identification, we use quaternions to compute all the amplitudes in the quantum geometric tensor. 
In the present Kramers setting, we choose a local occupied frame $(|u_1\rangle,|u_2\rangle)$ such that $|u_2\rangle=\Theta|u_1\rangle$.
The two complex states in this Kramers pair can be identified with a
single quaternionic state.
The Pauli matrix action on the doublet is identified with a quaternion multiplication. 
We denote the three quaternion basis as $\mathbf j^a, a=1,2,3$. 
The explicit identification is
\begin{equation}
(|u_1\rangle,|u_2\rangle)=\Psi \leftrightarrow \psi_{\mathbb H}, \qquad 
\Psi(-i\sigma_a)\leftrightarrow \psi_{\mathbb H}\mathbf j^a, \qquad 
D_\mu\Psi\leftrightarrow D_\mu\psi_{\mathbb H}.
\label{eq_qidf}
\end{equation}
Here, $\psi_{\mathbb H}$ is a $(N/2)$-dimensional quaternion vector. 
The state $D_\mu\psi_{\mathbb H}$ is a quaternion vector living in the $(N/2-1)$-dimensional space perpendicular to $\psi_{\mathbb H}$.

Let $v=v^\mu\partial_\mu$ be real tangent vectors in $T_{\bm\lambda}M$, and define $D_v\equiv v^\mu D_\mu$.
The action of $Q$ (and similarly, also for the metric $g$ and curvature $F$) on four dimensional tangent vectors $v,v'\in TM$ is done by contracting the $\mu,\nu$ indices of $v,v'$. 
Under this convention, the quantum geometric tensor is written as 
\begin{equation}
Q(v,v')\equiv v^\mu v'^\nu Q_{\mu\nu}
=\langle D_v\psi_{\mathbb H}|D_{v'}\psi_{\mathbb H}\rangle 
=g(v,v')+\sum_{a} F^a(v,v')\mathbf j^a, 
\qquad \textrm{Re}_{\mathbb H}\ Q(v,v')=g(v,v').\label{eq_qgth}
\end{equation}
Here, we rescale the normalization of $\psi_{\mathbb H}$ to $2$. The inner product of two quaternion vectors $q'_j,q_j$ is taken as $\sum_j\bar q_jq'_j$ with $\bar q$ the quaternion conjugation. With the quaternion algebra, we also have the relations $ \textrm{Re}_{\mathbb H}\ Q(v,v')\mathbf j^a=-F^a(v,v')$, paralleling the quantum geometric tensor decomposition in the $\Psi$ convention.

The quaternion Cauchy-Schwarz inequality tells that for any two quaternion vectors $q_j,q'_j$, we have $|\langle q|q'\rangle_{\mathbb H}|^2\le\langle q|q\rangle_{\mathbb H}\langle q'|q'\rangle_{\mathbb H}$. By setting $D_v\psi_{\mathbb H}=q$ and define $D_{v'}\psi_{\mathbb H}=q'$, we have 
\begin{equation}
|Q(v,v')|_{\mathbb H}\le \sqrt{Q(v,v)Q(v',v')}=\sqrt{g(v,v)g(v',v')},
\end{equation}
where the fact $Q(v,v)$ being real is employed. This yields the quaternion Cauchy-Schwarz inequality used in the main text for $\Delta_Q\ge 0$.

\subsection{Determinant bound saturation condition}
Then we prove the determinant bound saturation condition. According to the main text, the determinant bound is saturated when $[J^a]^\dagger J^a=1_{4\times 4}$ for all $a=1,2,3$. To demonstrate the quaternion Cauchy-Riemann condition, first, from definition we have $g(v,J^av')=F^a(v,v')$. On the other hand, we define the action $(D^\dagger q)^\mu=\sum_\nu g^{\mu\nu}\textrm{Re}_{\mathbb H}\langle D_v\psi_{\mathbb H}|q\rangle$ mapping a quaternion vector back to a tangent vector of the Brillouin zone. We use the conjugation symbol because $D$ maps a tangent vector $v$ to a quaternion vector $D_v\psi_{\mathbb H}$, while $D^\dagger$ acts in the reverse direction. Then we have 
\begin{equation}
g(v,D^\dagger[D_{v'}\psi_{\mathbb H}\mathbf j^a])=\textrm{Re}_{\mathbb H}\langle D_{v}\psi_{\mathbb H}|D_{v'}\psi_{\mathbb H}\mathbf j^a\rangle=-F^a(v,v').
\end{equation}
The first step comes from the definition of $D^\dagger$ and $\sum g_{\mu\rho}g^{\rho\nu}=\delta_\mu^\nu$, and the second step comes from the quantum geometric tensor decomposition Eq.~\eqref{eq_qgth}. Comparing the definition of $J^a$, we obtain 
\begin{equation}
[J^a]^\mu_\nu=-[D^\dagger (D_\nu\psi_{\mathbb H}\mathbf j^a)]^\mu.\label{eq_JJequiv}
\end{equation}

Now we are ready to derive the saturation condition. The key idea is to interpret the condition $1-[J^a]^\dagger J^a=0$ as the norm of a null vector, similar to the ideal conditions for two-dimensional Chern bands \cite{roy2014band}. We introduce the abbreviation $D$ for its action on a tangent vector ($Dv\equiv D_v\psi_{\mathbb H}$) and the notion $R^aq=-q\mathbf j^a$. The relation Eq.~\eqref{eq_JJequiv} is rewritten as $J^a=D^\dagger R^aD$. The saturation condition $1-[J^a]^\dagger J^a=0$ is translated into $1-D^\dagger[R^a]^\dagger DD^\dagger R^a D=0$, where $[R^a]^\dagger=-R^a$ is defined from the requirement $\textrm{Re}_{\mathbb H}\langle R^\dagger q'|q\rangle=\textrm{Re}_{\mathbb H}\langle q'|R^a q\rangle $. Note that $[R^a]^\dagger R^a=1$ and $D^\dagger D=1_{4\times 4}$ (obtained by inserting the definition of $D^\dagger$), we reduce the saturation condition to the form
\begin{equation}
1-[J^a]^\dagger J^a=0\ \Leftrightarrow\ D^\dagger[R^a]^\dagger(1-DD^\dagger)R^aD=0.
\end{equation}
Again using the relation $D^\dagger D=1$, the middle factor $(1-DD^\dagger)$ can be replaced by its square $(1-DD^\dagger)^2$. Plugging in this requirement into the metric, we obtain
\begin{equation}
 \Delta_Q=0 \Leftrightarrow\ g\big(v,(1-[J^a]^\dagger J^a)v\big)=0\ \forall v\ \Leftrightarrow \ \langle(1-DD^\dagger)R^aDv|(1-DD^\dagger)R^aDv\rangle=0\ \forall v.
\end{equation}
The last equation just tells $(1-DD^\dagger)R^aDv=0$. By inserting back the definitions and letting $v$ range over all four basis tangent vectors of $M$, the quaternion Cauchy-Riemann equations appear
\begin{equation}
D_\mu\psi_{\mathbb H}\mathbf j^a+\sum_\nu [J^a]^\nu_\mu D_\nu\psi_{\mathbb H}=0\ \leftrightarrow\  D_\mu\Psi(-i\sigma^a)+\sum_\nu [J^a]^\nu_\mu D_\nu\Psi=0.
\end{equation}
The second equation is obtained from the first through the identification  Eq.~\eqref{eq_qidf}. These equations can be conveniently written as $R^aD=DJ^a$, from which we have $R^aR^bD=DJ^aJ^b$. Applying $D^\dagger D=1$, we find that $J^a$ also satisfies the quaternion algebra $J^aJ^b=\sum_c \epsilon_{abc}J^c-\delta_{ab}$. This finishes the proof for the triple quaternion Cauchy-Riemann equations.

Throughout the proof, the key step is the right multiplication of $\mathbf j^a$ on the derivatives of the doublet $D_\mu\psi_{\mathbb H}$, or equivalently, the right multiplication by $(-i\sigma^a)$ on $D_\mu\Psi$. This means that the vector space where $D_\mu\Psi(\bm{\lambda)}$ takes values from at each $\bm\lambda$ is equipped with three quaternion multiplications. In the presence of $\Theta^2=-1$ symmetry, the existence of such quaternion operations is automatically guaranteed, because the sub-Hilbert spaces corresponding to the doublet form the quaternion projective space $\mathbb HP^{N/2-1}$. It is a quaternionic K\"ahler manifold whose tangent spaces, the vector spaces where $D_\mu\psi_{\mathbb H}$ reside, are equipped with three quaternion operations. On the other hand, if we remove the $\Theta^2=-1$ symmetry, the right multiplication by $(-i\sigma^a)$ still exists. The only information we need is that $D_\mu\Psi(-i\sigma^a)$ is perpendicular to $\Psi$ (being tangent), a local quaternion structure at each $\bm{\lambda}$. This does not require the whole Hilbert space of physical states $|u\rangle$ to be quaternionic, but only demanding locally for $D_\mu\Psi$ at each $\bm{\lambda}$. In fact, this local quaternion structure is indeed persists after removing the $\Theta^2=-1$ symmetry. In this case, the two-(complex) dimensional Hilbert space spanned by $\{|u_1\rangle,|u_2\rangle\}$ is characterized by the complex Grassmannian manifold $\textrm{Gr}_2(\mathbb C^2)$. It is also a quaternionic K\"ahler manifold whose tangent spaces are locally equipped with three quaternion operations \cite{wolfSpace,salamon1982quaternionic}. All of the above proofs can proceed by carefully seperating out the real part of the quantum geometric tensor and applying actions of Pauli matrices. In the next section, we give the details and apply the proof to a generic $U(2)$ doublet without additional symmetry protection.

\section{Inequality for $U(2)$ doubly degenerate bands}

In this section, we derive the local geometric bound for a general rank-two occupied subspace without assuming a $\Theta^2=-1$ symmetry. 
The essential ingredients are the positivity of a real Hilbert--Schmidt Gram matrix and the Pauli algebra associated with the rank-two occupied subspace.
In particular, no quaternionic structure of the full Hilbert space is required.
The notation is the same as the main text
\begin{equation}
\Psi=(|u_1\rangle,|u_2\rangle), \qquad
P=\Psi\Psi^\dagger, \qquad
X_\mu =D_\mu\Psi
=(1-P)\partial_\mu\Psi.
\end{equation}
The non-Abelian matrix-valued QGT is the matrix-valued Gram kernel of the interband amplitudes,
\begin{equation}
Q_{\mu\nu} =X_{\mu}^{\dagger} X_{\nu}
=G_{\mu\nu} -\frac{\ii}{2}F_{\mu\nu},\qquad
G_{\mu\nu}=\frac{1}{2} \big(Q_{\mu\nu} +Q_{\nu\mu} \big),\qquad
F_{\mu\nu}=\ii \big(Q_{\mu\nu} -Q_{\nu\mu} \big)
\label{eq:definition2}
\end{equation}
whose symmetric part gives the quantum metric $G_{\mu\nu}$ and the antisymmetric part gives the non-Abelian Berry curvature $F_{\mu\nu}$.
For any $U(2)$ curvature, we can separate the $U(1)$ and $SU(2)$ part as follows,
\begin{align}
F=\frac{1}{2}(\Tr F)I_2 + F^a\sigma_a,\qquad
{F}^a\equiv \frac{1}{2} F^a_{\mu\nu} d\lambda^{\mu}\wedge d\lambda^{\nu}
\end{align}
Similarly, for the quantum metric matrix, we can also separate it into the $U(1)$ and $SU(2)$ part,
\begin{align}
G_{\mu\nu}=\frac{1}{2} g_{\mu\nu} I_2 + G^a_{\mu\nu} \sigma_a,\qquad
g_{\mu\nu} =\Tr G_{\mu\nu}
\end{align}
with the scalar quantum metric $g_{\mu\nu}$.
In this way, we can have
\begin{align}
Q_{\mu\nu} =&X_{\mu}^{\dagger} X_{\nu}
=G_{\mu\nu} -\frac{\ii}{2}F_{\mu\nu}
=\Big(\frac{1}{2} g_{\mu\nu} -\frac{\ii}{4} (\Tr F_{\mu\nu}) \Big) I_2 
+ \Big( G^a_{\mu\nu} -\frac{\ii}{2} F_{\mu\nu}^a \Big)  \sigma_a
\end{align}
From this decomposition, for generic $U(2)$ case, the quantum geometric tensor is biquaternionic.
In the presence of a local Kramers symmetry $\Theta^2=-1$, 
the additional reality conditions $G^a=0$ and $\Tr F=0$ reduce the biquaternionic QGT to an ordinary quaternion-valued one.
Thus, the scalar quantum metric is just the real part of the trace over $Q$,
\begin{align*}
g_{\mu\nu} =\re \Tr Q_{\mu\nu}
=\re \Tr X_{\mu}^{\dagger} X_{\nu}
\end{align*}
Assuming the $4\times 4$ matrix $g$ has nonnegative eigenvalues $f_i$ ($i=1,\cdots,4$).
The arithmetic-geometric mean inequality gives
\begin{equation}
\frac14\sum_{i=1}^4 f_i
\geq(f_1 f_2 f_3 f_4)^{1/4},\qquad
\frac1{16}(\operatorname{tr} g)^2
\geq\sqrt{\det(g)}
\end{equation}
where $\operatorname{tr}$ is the trace over the parameter space index $\mu$.
Equality holds at a regular point if and only if the metric is isotropic, $g=\gamma I_4$, with a positive $\gamma>0$.

\subsection{Proof without assumptions of $\Theta^2=-1$ symmetry}
For a tangent $v^{\mu}$, we can introduce the corresponding $X_v$,
\begin{equation}
X_v=D_v\Psi =(1-P)\partial_v\Psi
=(1-P) \sum_{\mu} v^{\mu} \partial_{\mu}\Psi
=\sum_{\mu} v^{\mu} (1-P) \partial_{\mu}\Psi
=\sum_\mu v^\mu X_\mu.
\end{equation}
Since the occupied subspace has complex dimension two, $X_v$ can be regarded as a complex linear map
\begin{equation}
X_v:\mathbb C_{\mathrm{occ}}^2
\longrightarrow
\mathbb C_{\mathrm{emp}}^{N-2},
\end{equation}
or equivalently as an $(N-2)\times 2$ complex matrix $\mathbb C^{(N-2)\times 2}$.
We equip this space of $X_\mu$ with a real Hilbert-Schmidt inner product. For every two matrices $Y,Z\in \mathbb C^{(N-2)\times 2}$, the inner product is defined as
\begin{equation}
(Y,Z)_{\mathbb R} \equiv
\re\Tr(Y^\dagger Z),\qquad
g(v,w)= (X_v,X_w)_{\mathbb R},\qquad
g_{\mu\nu} =\re \Tr X_{\mu}^{\dagger} X_{\nu} 
=(X_{\mu},X_{\nu})_{\mathbb R},
\label{eq:real-HS}
\end{equation}
where the scalar quantum metric is the Gram metric induced by this inner product. This is actually the Riemannian metric on the complex Grassmanian $\textrm{Gr}_2(\mathbb C^N)$.
The non-Abelian Berry curvature for a general $U(2)$ occupied subspace can be decomposed as
\begin{equation}
F(v,w)=i\big(X_v^\dagger X_w-X_w^\dagger X_v\big)
=f(v,w)\mathbf 1_2 +\sum_{a=1}^{3} F^a(v,w)\sigma_a,
\label{eq:U2-decomposition}
\end{equation}
where the Pauli sector part is defined as,
\begin{equation}
F^a(v,w) =\frac{1}{2}
\Tr\big[ \sigma_a F(v,w)\big].
\end{equation}
We now show that the metric and the traceless curvature components are controlled by the positivity of a single Gram matrix.

For any fixed tangent vector $w$ of $M$, we introduce four related vectors in the space of $(N-2)\times 2$ complex matrices:
\begin{equation}
Y_0=X_w, \qquad
Y_a=X_w(-i\sigma_a), \qquad
a=1,2,3.
\label{eq:YA-definition}
\end{equation}
We first show that these four vectors are mutually orthogonal and have equal norms. 
Using the Hermiticity of the quantum geometric tensor, we have 
\begin{equation}
Q_{ww}=X_w^\dagger X_w=Q^{\dagger}_{ww},
\qquad\Rightarrow\qquad
\Tr(Q_{ww}\sigma_a)\in\mathbb R,\quad
\Tr Q_{ww}=\Tr(X_w^\dagger X_w) =g(w,w),
\end{equation}
Therefore, for $Y_0$ and $Y_a$ ($a,b=1,2,3$), we obtain
\begin{align}
(Y_0,Y_a)_{\mathbb R}
&=\re\Tr(Y_0^{\dagger}Y_a) =\re\Tr\big[X_w^{\dagger}X_w(-i\sigma_a)\big]
=\re\Tr\big[Q_{ww}(-i\sigma_a)\big]=0.    \\
(Y_a,Y_b)_{\mathbb R}
&=\re\Tr\left[(i\sigma_a)Q_{ww}(-i\sigma_b)\right]
=\re\Tr(\sigma_a Q_{ww}\sigma_b)
=\frac{1}{2}\Tr\left[Q_{ww}\{\sigma_a,\sigma_b\}\right]
=\delta_{ab}\Tr Q_{ww}
\end{align}
where we have used the relation $\re\Tr M=\frac{1}{2}(\Tr M+\Tr M^{\dagger})$ in both lines and the Pauli algebra $\{\sigma_a,\sigma_b\}=2\delta_{ab}\mathbf 1_2$ in the second line.
Together, the four vectors $Y_0,Y_a$ satisfy
\begin{equation}
(Y_A,Y_B)_{\mathbb R}
=g(w,w)\delta_{AB}, \qquad
A,B=0,1,2,3.
\label{eq:YA-orthogonality}
\end{equation}
Here, orthogonality holds as the real Hilbert-Schmidt inner product product is chosen (Riemannian metric), 
while in general in their complex overlaps are not vanishing.

Next, we compute the inner products between $Y_A$ and other vectors $X_v$. According to the definition Eq.~\eqref{eq:real-HS}, we have
\begin{align}
(X_v,Y_0)_{\mathbb R} =g(v,w),\qquad
(X_v,Y_a)_{\mathbb R}=\re\Tr
\big[X_v^\dagger X_w(-i\sigma_a) \big].
\end{align}
Defining $z_a=\Tr(X_v^\dagger X_w\sigma_a)$, we can have
\begin{align}
(X_v,Y_a)_{\mathbb R} =\operatorname{Im}z_a,\qquad
F^a(v,w)=\frac{i}{2}\Tr\left[\sigma_a
\left(X_v^\dagger X_w-X_w^\dagger X_v\right)\right]
=-\operatorname{Im}z_a.
\end{align}
Hence the inner products between $X_v$ and $Y_a$ are 
\begin{equation}
(X_v,Y_a)_{\mathbb R} =-F^a(v,w).
\label{eq:XvYa}
\end{equation}

Consider now the five vectors
\begin{equation}
X_v,\quad Y_0,\quad Y_1,\quad Y_2,\quad Y_3.
\end{equation}
Their Gram matrix with respect to the real Hilbert--Schmidt inner product is
\begin{equation}
\Gamma(v,w)=
\begin{pmatrix}
g(v,v) & g(v,w) & -F^1(v,w) & -F^2(v,w) & -F^3(v,w) \\
g(v,w) & g(w,w) & 0 & 0 & 0 \\
-F^1(v,w) & 0 & g(w,w) & 0 & 0 \\
-F^2(v,w) & 0 & 0 & g(w,w) & 0 \\
-F^3(v,w) & 0 & 0 & 0 & g(w,w)
\end{pmatrix}.
\label{eq:Gram-U2}
\end{equation}
Since $\Gamma(v,w)$ is a Gram matrix, it is positive semidefinite
$\Gamma(v,w)\geq 0$.
This indicates the following relation
\begin{gather}
\det\Gamma(v,w)=[g(w,w)]^3 
\left[g(v,v)g(w,w)- g(v,w)^2
- \sum_{a=1}^{3} [F^a(v,w)]^2 \right]
\geq 0. \\
g(v,v)g(w,w) -g(v,w)^2 
-\sum_{a=1}^{3}[F^a(v,w)]^2
\geq 0.
\label{eq:pairwise-U2-bound}
\end{gather}
The second equation also holds when $g(w,w)=0$. In that case, $X_w=0$ as the inner product is positively definite. Consequently we have $g(v,w)=F^a(v,w)=0$. Eq.~\eqref{eq:pairwise-U2-bound} holds trivially when $g(w,w)=0$.

Equation~\eqref{eq:pairwise-U2-bound} is the fundamental local inequality.
It is the direct analogue of the quaternion Cauchy-Schwarz inequality, but its derivation requires neither a $\Theta^2=-1$ symmetry nor a quaternionic structure on the full Hilbert space. 
The only special algebraic input is the Pauli-matrix identity $\{\sigma_a,\sigma_b\}= 2\delta_{ab}\mathbf1_2$,
which follows from the fact that the occupied subspace has complex dimension two. These relations originate from the \emph{local} quaternion structure of the tangent spaces $\mathrm{Gr}_2(\mathbb C^N)$, where all vectors $X_w$ reside.

We now specialize to a four-dimensional parameter space. 
At a point where the scalar quantum metric is positive definite, choose a $g$-orthonormal basis $\{e_1,e_2,e_3,e_4\}$, such that $g(e_\mu,e_\nu)=\delta_{\mu\nu}$.
For $\mu\neq\nu$, Eq.~\eqref{eq:pairwise-U2-bound} gives
\begin{align*}
\mu\neq\nu:\qquad \sum_{a=1}^{3} (F^a_{\mu\nu})^2 \leq 1
\end{align*}
Summing over the six independent two-planes, $(12),(13),(14),(23),(24),(34)$, we obtain
\begin{equation}
\sum_{a=1}^{3}\|F^a\|_g^2 \leq 6,\qquad 
\|F^a\|_g^2 = \frac12
F^a_{\mu\nu}F^{a\,\mu\nu}.
\end{equation}
For each real two-form $F^a$,
\begin{equation}
F^a\wedge F^a
=\langle F^a,*_gF^a\rangle_g\,\sqrt{\det g}\ d^4\lambda.
\end{equation}
Since the Hodge star is an isometry on two-forms,
\begin{equation}
\|*_gF^a\|_g=\|F^a\|_g,
\end{equation}
the ordinary Cauchy-Schwarz inequality gives
\begin{equation}
\left| \langle F^a,*_gF^a\rangle_g \right| 
\leq \|F^a\|_g^2.
\end{equation}
Therefore,
\begin{align}
\left| \sum_{a=1}^{3} F^a\wedge F^a \right|
&\leq \sum_{a=1}^{3} \|F^a\|_g^2\,\sqrt{\det g}
\leq 6\,\sqrt{\det g}.
\end{align}
Hence the local determinant bound is
\begin{equation}
\sqrt{\det g} \geq \frac16
\left| \sum_{a=1}^{3} F^a\wedge F^a \right|.
\label{eq:U2-volume-bound}
\end{equation}

Finally, the bound can be expressed in terms of the differential forms carrying Chern classes. 
Using the relations
\begin{gather}
F=f\mathbf1_2+ \sum_{a=1}^{3}F^a\sigma_a,\qquad
c_1=\frac{\Tr F}{2\pi},\qquad
c_2=\frac{ \Tr(F\wedge F)-(\Tr F)\wedge(\Tr F)}{8\pi^2},    \\
\Tr F=2f,\qquad
\Tr(F\wedge F) =2f\wedge f
+2\sum_{a=1}^{3} F^a\wedge F^a,
\end{gather}
one finds
\begin{equation}
c_2 + \frac14c_1\wedge c_1 = \frac{1}{4\pi^2}
\sum_{a=1}^{3} F^a\wedge F^a.
\label{eq:Chern-traceless}
\end{equation}
Substituting Eq.~\eqref{eq:Chern-traceless} into
Eq.~\eqref{eq:U2-volume-bound}, we obtain
\begin{equation}
U(2):\qquad \sqrt{\det g}\geq \frac{2\pi^2}{3}
\left| c_2+\frac14c_1\wedge c_1 \right|.
\label{eq:general-U2-bound}
\end{equation}
For an $SU(2)$ occupied bundle, $\Tr F=0$ and hence $c_1=0$. 
The general $U(2)$ result therefore reduces immediately to
\begin{equation}
SU(2):\qquad 
\sqrt{\det g}\geq \frac{2\pi^2}{3}|c_2|.
\end{equation}
The derivation shows that the geometric inequality is a consequence of Gram-matrix positivity. 
The special role of rank two enters through the Pauli algebra, which ensures that, for every $X_w$, the four vectors
\begin{equation}
X_w,\qquad
X_w(-i\sigma_1),\qquad
X_w(-i\sigma_2),\qquad
X_w(-i\sigma_3)
\end{equation}
form an orthogonal four-dimensional real subspace with equal norms. 
The present Gram-matrix proof based on Gram-matrix positivity actually corresponds to the quaternionic Cauchy-Schwarz inequality in the $\Theta^2=-1$ formulation if we treat $(-i\sigma^a)$ as the local quaternion structure on the tangent spaces of $\textrm{Gr}_2(\mathbb C^N)$.

\subsection{Saturation condition for a general rank-two $U(2)$ occupied subspace}

We now derive the saturation condition for the general rank-two $U(2)$ geometric bound. 
The derivation closely parallels the quaternionic Cauchy-Schwarz proof, without explicitly invoking the quaternionic structure of the full Hilbert space. 
Instead, the quaternionic action used below acts only on the occupied-band index of the interband amplitudes.
At a point where the scalar quantum metric is positive definite, we regard the space of interband amplitudes as a Hilbert space equipped with the real Hilbert--Schmidt inner product
\begin{equation}
T \Psi=\operatorname{Hom}_{\mathbb C}
\left(\mathbb C_{\mathrm{occ}}^2, \mathbb C^{N-2}_{\mathrm{emp}} \right), \quad (Y,Z)_{\mathbb R}=
\operatorname{Re}\Tr(Y^\dagger Z),\ \forall Y,Z\in T\Psi.
\end{equation}
where $T\Psi$ represents the fact that all $D_v\Psi$ vectors are perpendicular to $\Psi$, tangent to the variations of $\Psi$. 
In fact,  this is the tangent space of the complex Grassmannian $\mathrm{Gr}_2(\mathbb C^N)$ that represents two-dimensional subspaces of $\mathbb C^N$.
We introduce the linear map
\begin{equation}
D:T_{\lambda}M\longrightarrow T\Psi,
\qquad Dv\equiv X_v=D_v\Psi ,\qquad
g(v,w)=(Dv,Dw)_{\mathbb R}.
\label{eq:U2-D-definition}
\end{equation}
The adjoint $D^\dagger:T\Psi\rightarrow T_{\lambda}M$
is defined by
\begin{equation}
g(v,D^\dagger Y)
=(Dv,Y)_{\mathbb R}.
\label{eq:U2-Ddagger-definition}
\end{equation}
From this definition, direct calculations immediately gives
\begin{equation}
D^\dagger D=1_{T_{\lambda}M}.
\label{eq:U2-DdaggerD}
\end{equation}
Therefore, $DD^\dagger$ is the orthogonal projector in $T\Psi$ onto the image of $D$ (this can be verified by acting $DD^\dagger$ on any $D_v\Psi$), 
a four-dimensional (real dimension) subspace of $T\Psi$.

We proceed the proof with Pauli matrices. 
As before, we introduced the right action on the interband-amplitude space $T\Psi$, defined by
\begin{equation}
R^aX
=X(i\sigma_a)
\label{eq:U2-R-definition}
\end{equation}
With respect to the real Hilbert-Schmidt inner product, the right actions satisfy
\begin{equation}
[R^a]^\dagger=-R^a,\qquad
[R^a]^\dagger R^a=1,\qquad
R^aR^b=\epsilon_{abc}R^c-\delta_{ab}.
\label{eq:U2-R-adjoint}
\end{equation}
In analogy with the quaternionic formulation, we define three endomorphisms of the tangent space by
\begin{equation}
J^a \equiv D^\dagger R^aD.
\label{eq:U2-J-definition}
\end{equation}
Using the definition of $R^a$ and the relation
$(X_v,X_w(-i\sigma_a))_{\mathbb R}=-F^a(v,w)$ from Eq.~\eqref{eq:XvYa},
we find the relations
\begin{align}
g(v,J^aw)&=(Dv,R^aDw)_{\mathbb R}
=(Dv,Dw(i\sigma_a))_{\mathbb R}
=F^a(v,w).
\label{eq:U2-F-J-relation}
\end{align}
Thus, $J^a$ is the tangent-space endomorphism associated with the traceless Berry-curvature component $F^a$. 
The conjugation of $J^a$ is as defined in the main text. 
From the property of the real inner product, it satisfies
\begin{equation}
[J^a]^\dagger
=D^\dagger[R^a]^\dagger D
=-J^a.
\label{eq:U2-J-antisymmetric}
\end{equation}

We now rewrite the saturation condition in operator forms. 
From Eqs.~\eqref{eq:U2-DdaggerD}, \eqref{eq:U2-R-adjoint}, and
\eqref{eq:U2-J-definition}, one obtains
\begin{align}
1-[J^a]^\dagger J^a
&=D^\dagger D -D^\dagger[R^a]^\dagger D D^\dagger R^aD
=D^\dagger[R^a]^\dagger (1-DD^\dagger) R^aD.
\label{eq:U2-J-defect}
\end{align}
Since $DD^\dagger$ is an orthogonal projector,
$1-DD^\dagger$ is also an orthogonal projector. Therefore, for every tangent vector $v$,
\begin{align}
g\left(v,\left[1-[J^a]^\dagger J^a\right]v\right)
=\left(R^aDv,(1-DD^\dagger)R^aDv\right)_{\mathbb R}
=\left\|(1-DD^\dagger)R^aDv\right\|_{\mathbb R}^2
\geq 0
\label{eq:U2-null-vector-norm}
\end{align}
Hence, we recover the result $1-[J^a]^\dagger J^a\geq0.$
The curvature norm can be expressed directly in terms of $J^a$.
In a $g$-orthonormal tangent frame, Eq.~\eqref{eq:U2-F-J-relation} leads to
\begin{equation}
\|F^a\|_g^2=\frac12 \operatorname{tr}\left([J^a]^\dagger J^a\right),
\label{eq:U2-curvature-J-norm}
\end{equation}
where $\mathrm{tr}$ denotes the trace over the four-dimensional tangent space of $M$. 
We therefore define the Gram defect
\begin{equation}
\Delta_Q\equiv6-\sum_{a=1}^{3}\|F^a\|_g^2
=\frac12\sum_{a=1}^{3}\operatorname{tr}\left(1-[J^a]^\dagger J^a\right).
\label{eq:U2-DeltaQ}
\end{equation}
Since each operator $1-[J^a]^\dagger J^a$ is positive semidefinite,
the saturation condition $\Delta_Q=0$ requires each of them to vanish separately:
\begin{equation}
\Delta_Q=0 \quad\Longleftrightarrow\quad
1-[J^a]^\dagger J^a=0, \qquad a=1,2,3.
\label{eq:U2-DeltaQ-J-saturation}
\end{equation}
Using Eq.~\eqref{eq:U2-null-vector-norm}, the latter condition can be written as a null-vector condition,
\begin{align}
\Delta_Q=0 &\Longleftrightarrow
g\left(v,\left[1-[J^a]^\dagger J^a\right]v \right)=0
\qquad \forall v,\quad a=1,2,3 \nonumber\\
&\Longleftrightarrow
\left\|(1-DD^\dagger)R^aDv\right\|_{\mathbb R}^2=0
\qquad \forall v,\quad a=1,2,3 \nonumber\\
&\Longleftrightarrow
(1-DD^\dagger)R^aDv=0
\qquad \forall v,\quad a=1,2,3.
\label{eq:U2-null-condition}
\end{align}
The last equation has a simple geometric interpretation:
the four-dimensional image $\operatorname{Img}D$ of the operator $D$ in $T\Psi$ must be invariant under all three right quaternionic actions $R^a$,
\begin{equation}
R^a(\operatorname{Img}D)
\subseteq \operatorname{Img}D,
\qquad a=1,2,3.
\label{eq:U2-invariant-subspace}
\end{equation}
Using $DD^\dagger$ as the projector to $\operatorname{Img}D$, the condition~\eqref{eq:U2-null-condition} gives
\begin{align}
R^aD=DD^\dagger R^aD =D J^a.
\end{align}
Hence, the saturation condition can be written compactly as
\begin{equation}
R^aD=DJ^a, \qquad a=1,2,3.
\label{eq:U2-intertwining}
\end{equation}
Applying this equation to the basis tangent vector $\partial_\mu$,
and using $R^aX=-X(-i\sigma^a)$, we obtain
\begin{gather}
-D_\mu\Psi\,(-i\sigma^a)
= \sum_{\nu} [J^a]^\nu{}_\mu D_\nu\Psi,  \\
D_\mu\Psi\,(-i\sigma^a)
+\sum_{\nu}[J^a]^\nu{}_\mu
D_\nu\Psi =0,
\qquad a=1,2,3.
\label{eq:U2-triple-CR}
\end{gather}
These are the triple quaternionic Cauchy--Riemann equations for a general rank-two $U(2)$ occupied subspace.

The quaternion algebra of the tangent-space endomorphisms $J^a$ follows directly from the intertwining relation Eq.~\eqref{eq:U2-intertwining}. 
Acting twice, we have
\begin{equation}
R^aR^bD =DJ^aJ^b.
\end{equation}
On the other hand, 
\begin{equation}
R^aR^bD=\left(\epsilon_{abc}R^c-\delta_{ab}\right)D
=D\left(\epsilon_{abc}J^c-\delta_{ab}\right).
\end{equation}
Since $D^\dagger D=1$, the map $D$ is injective, and therefore
\begin{equation}
J^aJ^b =\epsilon_{abc}J^c-\delta_{ab}.
\label{eq:U2-J-quaternion-algebra}
\end{equation}
In particular,
\begin{equation}
(J^a)^2=-1, \qquad [J^a]^\dagger J^a=1.
\end{equation}
Thus, although no quaternionic structure of the full Hilbert space was assumed, saturation forces the tangent image $\operatorname{Im}D$ to become a quaternionic invariant four-dimensional real subspace.

We finally relate this condition to the saturation of the four-dimensional volume bound. 
Choose a unit tangent vector $e_1$ and introduce the orthonormal frame
\begin{equation}
e_2=J^1e_1,\qquad
e_3=J^2e_1,\qquad
e_4=J^3e_1.
\label{eq:U2-adapted-frame}
\end{equation}
Let $\{\xi^1,\xi^2,\xi^3,\xi^4\}$ be the corresponding dual coframe.
Using
\begin{equation}
F^a(v,w)=g(v,J^aw)
\end{equation}
and the quaternion algebra Eq.~\eqref{eq:U2-J-quaternion-algebra}, the three curvature two-forms take the canonical form
\begin{align}
F^1&=-\left( \xi^1 \wedge \xi^2
+\xi^3 \wedge \xi^4 \right), \nonumber\\
F^2&=-\left( \xi^1 \wedge \xi^3
- \xi^2 \wedge \xi^4 \right), \nonumber\\
F^3 &=-\left( \xi^1 \wedge \xi^4
+\xi^2 \wedge \xi^3 \right),
\label{eq:U2-canonical-curvature}
\end{align}
up to a common orientation and an $SO(3)$ rotation of the Pauli indices. 
Consequently, with respect to the chosen orientation, there exists a common sign $s=\pm1$ such that
\begin{equation}
*_gF^a=sF^a,\qquad a=1,2,3.
\label{eq:U2-common-self-duality}
\end{equation}
Moreover,
\begin{equation}
\|F^a\|_g^2=2,\qquad F^a\wedge F^b
=2s\,\delta_{ab}\,\sqrt{\det g},\qquad
\sum_{a=1}^{3}\|F^a\|_g^2=6,\qquad
\sum_{a=1}^{3}F^a\wedge F^a
=6s\,\sqrt{\det g}.
\label{eq:U2-F-wedge-canonical}
\end{equation}
Therefore,
\begin{equation}
\sqrt{\det g}=\frac16
\left|\sum_{a=1}^{3}F^a\wedge F^a\right|.
\label{eq:U2-volume-saturation}
\end{equation}

Conversely, saturation of the volume bound
\begin{equation}
\sqrt{\det g}\geq\frac16
\left|\sum_{a=1}^{3}F^a\wedge F^a\right|,\qquad
\left|\sum_aF^a\wedge F^a\right|
\leq\sum_a\|F^a\|_g^2\,\sqrt{\det g}
\leq 6\,\sqrt{\det g}.
\end{equation}
requires the condition
\begin{equation}
\sum_a\|F^a\|_g^2=6,
\end{equation}
or equivalently $\Delta_Q=0$. Hence, at every point where $g$ is positive definite,
\begin{equation}
\sqrt{\det g}
=\frac16\left|\sum_aF^a\wedge F^a\right|
\quad\Longleftrightarrow\quad
\Delta_Q=0
\quad\Longleftrightarrow\quad
R^aD=DJ^a \quad (a=1,2,3).
\label{eq:U2-saturation-equivalence}
\end{equation}
Equivalently, the saturation condition is given by the triple Cauchy-Riemann equations
\begin{equation}
D_\mu\Psi(-i\sigma_a)
+\sum_\nu[J^a]^\nu{}_\mu D_\nu\Psi
=0, \qquad a=1,2,3.
\end{equation}
Importantly, this saturation condition does not require $\Tr F=0$ or $G_{\mu\nu}^a=0$. 
The quaternionic structure appearing in Eqs.~\eqref{eq:U2-intertwining}-\eqref{eq:U2-J-quaternion-algebra} is instead an emergent structure of the saturated tangent image $\operatorname{Img}D$ inside $T\Psi$.

Below, we give a table in Tab.~(\ref{tab:compCP2HP1}) listing out the thorough comparison between the results for $SU(2)$ doublets and the results for $U(2)$ doublets.

\begin{table}[htbp]
\centering
{\renewcommand{\arraystretch}{1.9}
\begin{tabular}{c|c|c}
\hline\hline
\textbf{Property}
& \textbf{$\mathbb{CP}^{2}$ model}
& \textbf{$\mathbb{HP}^{1}\simeq S^{4}$ model}\\
\hline
\makecell{Representative Hamiltonian}
& $\begin{gathered}
H=2|v\rangle\langle v|-\mathbf{1}_{3},\\
P=\mathbf{1}_{3}-|v\rangle\langle v|
\end{gathered}$
& $\begin{gathered}
H=n_{A}\Gamma_{A},\quad n_{A}n_{A}=1,\\
P=(\mathbf{1}_{4}-H)/2
\end{gathered}$ 
\\
\hline
\makecell{Hilbert-space dimension }
& $\begin{gathered} 
N=3:\\ 2\text{ occupied}+1\text{ empty}
\end{gathered}$
& $\begin{gathered}
N=4:\\ 2\text{ occupied}+2\text{ empty}
\end{gathered}$ 
\\
\hline
\makecell{Parameter space}
& $\mathbb{CP}^{2}
\simeq\mathrm{Gr}_{2}(\mathbb{C}^{3})$
& $\mathbb{HP}^{1}\simeq S^{4}
\subset\mathrm{Gr}_{2}(\mathbb{C}^{4})$
\\
\hline
\makecell{Occupied bundle}
& $\begin{gathered}
\text{Universal rank-two}\\
\text{quotient bundle over }\mathbb{CP}^{2}
\end{gathered}$
& $\begin{gathered}
\text{Quaternionic line bundle,}\\
\text{viewed as complex rank two}
\end{gathered}$
\\
\hline
\makecell{Structure group}
& $U(2)$ ($f\neq0$)
& $Sp(1)\simeq SU(2)$ ($f=0$)
\\
\hline
\makecell{Kramers symmetry}
& $\begin{gathered}
\Theta^{2}=-1\text{ is impossible}\\
\text{in the minimal }N=3\text{ space}
\end{gathered}$
& $\begin{gathered}
\text{A standard Kramers or}\\
\text{quaternionic structure exists}
\end{gathered}$
\\
\hline
\multicolumn{3}{c}{\textbf{Characteristic classes}}
\\
\hline
\makecell{First Chern class}
& $c_{1}=h,\quad\displaystyle\int_{\mathbb{CP}^{2}}h^{2}=1$
& $c_{1}=0$
\\
\hline
\makecell{Second Chern class}
& $c_{2}=-h^{2}$
& $c_{2}=s u,\quad \displaystyle\int_{S^{4}}u=1,\quad s=\pm1$
\\
\hline
Adjoint four-form
& $\displaystyle \Xi=c_{2}+\frac14c_{1}^{2}=-\frac34h^{2}$
& $\Xi=\varepsilon u$
\\
\hline
\makecell{Mathematical \\ interpretation}
& $\displaystyle
\Xi=\frac14p_{1}(\mathfrak{su}E)$
& $SU(2)$ instanton density
\\
\hline
Integrated invariants
& $C_{1,1}=1,\quad C_{2}=-1$
& $C_{1,1}=0,\quad C_{2}=s$
\\
\hline
\multicolumn{3}{c}{\textbf{Metric-curvature}}
\\
\hline
\makecell{Adjoint determinant}
& \multicolumn{2}{c}{$\displaystyle\mathrm{vol}_{g}
=\frac16\Big|\sum_{a}F^{a}\wedge F^{a}\Big|
=\frac{2\pi^{2}}{3}|\Xi|$}
\\
\hline
\makecell{Abelian Wirtinger}
& $\displaystyle\mathrm{vol}_{g}=\frac12f\wedge f
=\frac{\pi^{2}}{2}c_{1}^{2}$
& Trivial because $f=0$
\\
\hline
\makecell{Integrated volume bound}
& $\displaystyle\operatorname{Vol}_{g}=\frac{\pi^{2}}{2}|C_{1,1}|$
& $\displaystyle\operatorname{Vol}_{g}=\frac{2\pi^{2}}{3}|C_{2}|$
\\
\hline
\multicolumn{3}{c}{\textbf{Saturation and quaternionic geometry}}
\\
\hline
Quantum defect
& \multicolumn{2}{c}{$\displaystyle\Delta_{Q}
=6-\sum_{a}\|F^{a}\|_{g}^{2}=0$} 
\\
\hline
Hodge condition
& \makecell{$\displaystyle*_{g}F^{a}=-F^{a}$ with \\ the complex orientation}
& $\displaystyle*_{g}F^{a}=s F^{a}$
\\
\hline
\makecell{Transition-amplitude \\ closure}
& $
T\mathbb{CP}^{2}\text{ is closed under }
X\mapsto X(-i\sigma_{a})$
& $TS^{4}\text{ is closed under }
X\mapsto X(-i\sigma_{a})$
\\
\hline
\multicolumn{3}{c}{\textbf{Metric and Yang--Mills properties}}
\\
\hline Quantum metric
& Fubini--Study metric
& $\displaystyle g_{\mu\nu}=\frac12\partial_{\mu}\boldsymbol{n}\cdot\partial_{\nu}\boldsymbol{n}$
\\
\hline
\makecell{Canonical \\ Einstein geometry}
& $R_{\mu\nu}=6g_{\mu\nu},\quad R=24$
& $\begin{gathered}
\text{Round }S^{4}\text{ of radius }1/\sqrt{2},\\
R_{\mu\nu}=6g_{\mu\nu},\quad R=24
\end{gathered}$
\\
\hline
\makecell{Curvature structure}
& $\begin{gathered}
f\text{ is self-dual},\\
F^{a}\text{ are anti-self-dual}
\end{gathered}$
& $F^{a}$ self or anti-self-dual
\\
\hline\hline
\end{tabular}
}
\caption{Comparison between the $\mathbb{CP}^{2}$ three-level model and the $\mathbb{HP}^{1}\simeq S^{4}$ Yang-monopole model.
We use $F=f\mathbf{1}_{2}+\sum_{a=1}^{3}F^{a}\sigma_{a}$, $c_{1}=\mathrm{Tr}F/(2\pi)$,  and $c_{2}=[\mathrm{Tr}(F\wedge F)-(\mathrm{Tr}F)\wedge(\mathrm{Tr}F)]/(8\pi^{2})$.}
\label{tab:compCP2HP1}
\end{table}

\subsection{The $\mathbb{C}P^2$ model: a minimal saturated $U(2)$ quantum geometry}
\label{app:cp2-model}
We now discuss a minimal three-level model which saturates the $U(2)$ geometric bound.  
Besides providing a concrete example of the $U(2)$ inequality, this model also illustrates an important distinction between the generic $U(2)$ geometry and the genuine $SU(2)$ geometry discussed in the main text. 
The comparison between the two model is summarized in Tab.~\ref{tab:compCP2HP1}.
Consider the following Hamiltonian constructed from a singlet state $|v\rangle\equiv |v(w_1,w_2)\rangle$,
\begin{align}
H=2|v\rangle\langle v|-I_3,\qquad
|v\rangle=\frac{(1,w_1,w_2)^T}
{\sqrt{1+|w_1|^2+|w_2|^2}},\qquad
w_1=\lambda^1+\ii\lambda^2,\qquad
w_2=\lambda^3+\ii\lambda^4.
\label{eq:cp2-app-hamiltonian}
\end{align}
Here, the normalized state $|v\rangle$ has energy $+1$, while its orthogonal complement is a two-dimensional degenerate eigenspace $\Psi=(|u_1,\rangle,|u_2\rangle)$ with energy $-1$.
The two occupied bands are separated from a single empty band, and the occupied projector is simply
\begin{align}
P=\Psi\Psi^{\dagger} 
=|u_1\rangle \langle u_1|
+|u_2\rangle \langle u_2|
=I_3-|v\rangle\langle v|
=\frac{1}{2} (I_3-H).
\label{eq:cp2-app-projector}
\end{align}
Since an overall phase of $|v\rangle$ leaves both $H$ and $P$ unchanged, the physical parameter space is the space of normalized three-component complex vectors modulo an overall phase, namely $M=S^5/U(1)=\mathbb{C}P^2$.
The coordinates $(w_1,w_2)$ cover one affine patch $\mathbb{C}^2\subset\mathbb{C}P^2$.

The interband matrix becomes
\begin{align}
X_\mu=(1-P)\partial_\mu\Psi
=|v\rangle\langle v|\partial_\mu\Psi,\qquad
(X_\mu)_{vn}
=\langle v|\partial_\mu u_n\rangle,
\qquad n=1,2.
\label{eq:cp2-app-X}
\end{align}
Therefore all virtual transitions generated by changing the four parameters are forced to pass through the same empty level. 
Indeed, $\varepsilon_n-\varepsilon_v=-2$, while
\begin{align}
\partial_\mu H
=2|\partial_\mu v\rangle\langle v|
+2|v\rangle\langle\partial_\mu v|,\qquad
\frac{\langle v|\partial_\mu H|u_n\rangle}
{\varepsilon_n-\varepsilon_v}
=-\langle\partial_\mu v|u_n\rangle
=\langle v|\partial_\mu u_n\rangle,
\end{align}
where in the last equality we used $\partial_\mu\langle v|u_n\rangle=0$.
The scalar quantum metric can be evaluated without introducing an explicit basis for the two occupied states.  
Using $\sum_{n=1}^{2}|u_n\rangle\langle u_n|=P$, one finds
\begin{align}
\Tr Q_{\mu\nu}&=\sum_{n=1}^{2}
\langle\partial_\mu u_n|v\rangle
\langle v|\partial_\nu u_n\rangle
=\langle\partial_\nu v|
P|\partial_\mu v\rangle ,\qquad
g_{\mu\nu}
=\re
\langle\partial_\mu v|
P|\partial_\nu v\rangle .
\label{eq:cp2-app-metric-projector}
\end{align}
Thus the scalar metric of the two occupied bands is identical to the Fubini-Study metric associated with the complementary state $|v\rangle$.
For the parametrization in Eq.~\eqref{eq:cp2-app-hamiltonian}, its line
element is
\begin{align}
g_{\mu\nu}d\lambda^\mu d\lambda^\nu
=\frac{\left(1+|w_1|^2+|w_2|^2\right)\left(|dw_1|^2+|dw_2|^2\right)
-\left|\bar w_1dw_1+\bar w_2dw_2\right|^2
}{\left(1+|w_1|^2+|w_2|^2\right)^2}.
\label{eq:cp2-app-fsmetric}
\end{align}
Since $w_1$ and $w_2$ contain four real coordinates, the determinant of the corresponding real $4\times4$ metric is therefore
\begin{align}
\det g=\frac{1}{\left(1+|w_1|^2+|w_2|^2\right)^6},\qquad
\sqrt{\det g}
=\frac{1}{
\left(1+|w_1|^2+|w_2|^2\right)^3}.
\label{eq:cp2-app-volume-density}
\end{align}

The Abelian part of the non-Abelian Berry curvature can be obtained from the same projector expression.  
With the convention $Q_{\mu\nu}=G_{\mu\nu}-\ii F_{\mu\nu}/2$, one obtains
\begin{align}
\Tr F=\frac{\ii}{
\left(1+|w_1|^2+|w_2|^2\right)^2}
\Big[&(1+|w_2|^2)\, dw_1\wedge d\bar w_1
+(1+|w_1|^2)\, dw_2\wedge d\bar w_2 \nonumber\\
& -\bar w_1w_2\, dw_1\wedge d\bar w_2
-\bar w_2w_1\, dw_2\wedge d\bar w_1 \Big].
\label{eq:cp2-app-trF}
\end{align}
Taking the wedge product with itself gives
\begin{align}
(\Tr F)\wedge(\Tr F)
=\frac{8\,d\lambda^1\wedge d\lambda^2
\wedge d\lambda^3\wedge d\lambda^4
}{\left(1+|w_1|^2+|w_2|^2\right)^3}.
\label{eq:cp2-app-trF2}
\end{align}
Combining Eqs.~\eqref{eq:cp2-app-volume-density} and
\eqref{eq:cp2-app-trF2}, we obtain the local equality
\begin{align}
c_1=\frac{\Tr F}{2\pi},\qquad
\sqrt{\det g}\, d^4 \lambda
=\frac12\left(\frac{\Tr F}{2}\right)
\wedge\left(\frac{\Tr F}{2}\right)
=\frac{\pi^2}{2}\, c_1\wedge c_1.
\label{eq:cp2-app-c1-saturation}
\end{align}
Here and below $c_1^2$ is understood as $c_1\wedge c_1$.
The affine coordinates $(w_1,w_2)$ cover $\mathbb{C}^2$,
while the omitted $\mathbb{C}P^1$ at infinity has zero measure with
respect to the four-dimensional volume form.  Therefore
\begin{align}
\operatorname{vol}_g(\mathbb{C}P^2)
&=\int_{\mathbb{C}^2}
\frac{d^4\lambda}{\left(1+|w_1|^2+|w_2|^2\right)^3}
=2\pi^2\int_0^\infty\frac{r^3\,dr}{(1+r^2)^3}
=\frac{\pi^2}{2}.
\end{align}
with  $|w_1|^2+|w_2|^2=r^2$ and using $d^4\lambda=2\pi^2r^3dr$.
Consequently, for the complex orientation used above,
\begin{gather}
C_{1,1}=\int_{\mathbb{C}P^2}c_1\wedge c_1=1,\qquad
C_2=\int_{\mathbb{C}P^2}c_2=-1, \\
\operatorname{vol}_g(\mathbb{C}P^2)
=\frac{\pi^2}{2}|C_{1,1}|
=\frac{\pi^2}{2}|C_2|
=\frac{\pi^2}{2}.
\label{eq:cp2-app-global-saturation}
\end{gather}

\end{document}